\documentclass[BCOR=10mm,DIV=calc,12pt]{scrartcl}
\KOMAoptions{DIV=calc}
\setkomafont{author}{\sffamily}

\usepackage{graphicx}
\usepackage{subcaption}
\usepackage{hyperref}
\usepackage{siunitx}
\usepackage{amssymb}
\usepackage{authblk}
\usepackage{booktabs}
\usepackage{makecell}
\usepackage{xcolor}
\usepackage{appendix}

\usepackage[colorinlistoftodos,prependcaption,textsize=tiny]{todonotes}

\usepackage[a4paper]{geometry}
\title{Resolution Enhancement of Scanning Electron Micrographs using Artificial Intelligence}
\author[1]{T. Reclik*}
\author[1]{S. Medghalchi}
\author[1]{P. Schumacher}
\author[1]{M. Wollenweber}
\author[1]{T. Al-Samman}
\author[1]{S. Korte-Kerzel}
\author[2]{U. Kerzel}
\affil[1]{Institute for Physical Metallurgy and Materials Physics, RWTH Aachen University, Aachen, Germany}
\affil[2]{Data Science and Artificial Intelligence in Materials and Geoscience, Fakultät für Georessourcen und Materialtechnik, RWTH Aachen University, Aachen, Germany}

\begin{document}
\maketitle

\begin{abstract}
\noindent Scanning Electron Microscopy (SEM) is pivotal in revealing intricate micro- and nanoscale features across various research fields. However, obtaining high-resolution SEM images presents challenges, including prolonged scanning durations and potential sample degradation due to extended electron beam exposure. This paper addresses these challenges by training and applying a deep learning based super-resolution algorithm. We show that the chosen algorithm is capable of increasing the resolution by a factor of 4, thereby reducing the initial imaging time by a factor of 16. We benchmark our method in terms of visual similarity and similarity metrics on two different materials, a dual-phase steel 
and a case-hardening steel, improving over standard interpolation methods. Additionally, we introduce an experimental pipeline for the study of rare events in scanning electron micrographs, without losing high-resolution information.
\par\vskip\baselineskip\noindent
\end{abstract}

\section{Introduction}

Scanning Electron Microscopy (SEM) offers a unique capability to probe features at the micro- and nanoscale levels, enabling detailed analyses of diverse samples and revealing their properties, such as composition and structure, and enabling the connection between physical mechanisms and changes in the microstructure through in-situ experiments. Such depth of information is been crucial for various fields, ranging from archaeology
\cite{pedergnana_new_2020,gleba_textiles_2012,cardell_innovative_2009}, to battery
research \cite{golozar_situ_2018,chen_situ_2016,liu_microstructural_2020}, and
materials science \cite{li_multislip-enabled_2023,mathis_dynamics_2021,alatarvas_revealing_2023}. 

Despite its enormous benefits, obtaining high-resolution scanning electron
micrographs is not without challenges. 
Time consumption is one of the key issues, with high-resolution
imaging requiring substantial scanning duration, especially for larger sample areas, e.g. the imaging of an area of \SI{1}{\milli\meter\squared} at a resolution of \SI{32.5}{\nano \meter} per pixel at a dwell time of \SI{32}{\micro\second} would take \SI{9}{\hour}.
Traditionally, many analyses are limited by the area that can be captured in
high-resolution images and are typically limited to areas of a magnitude of \SI{10}{\micro\meter} in edge length, but can sometimes be significantly larger. However, to be able to draw statistically 
relevant conclusions, large areas need to be analysed \cite{kusche_large-area_2019, medghalchi_three-dimensional_2023}. Especially rare events are often inhomogeneously distributed over large areas of a specimen, requiring to scan a substantial part of the sample in order to be able to 
find and potentially track these spots, for example when tracking damage sites over different deformation steps in in-situ tensile experiments \cite{wollenweber_automated_2023, qayyum_influence_2022, tasan_strain_2014}.
Furthermore, a long scan duration can additionally lead to 
imaging artefacts due to, for example, drift that, depending on the application, need to be corrected afterwards \cite{sutton_metrology_2006,sutton_scanning_2007,jin_correction_2015}.
Finally, long exposure to electron beams can induce alterations or even degradation
in the sample \cite{egerton_radiation_2004}. Mitigating this by limiting exposure time to the electron beam, however, results in poorer image quality.

Ideally, we would like to take high-resolution images, while keeping recording times short. As this is not possible, alternatively it is possible to record in a lower resolution and improve the resolution in post. Simple methods, such as,
bi-cubic interpolation, enable us to improve the resolution to some extent. However, these 
methods are not powerful enough for most scientific analyses. More sophisticated methods, termed super-resolution in the realm of computer vision, learn to infer high-resolution images from their lower-resolution counterparts using machine learning. 
Conventionally, in the realm of  light microscopy, 
the term ``super-resolution'' refers to a set of techniques
that surpass the diffraction limit of light, allowing the visualisation of features smaller than the
wavelength of light, for example, using methods such as STED\cite{hell_breaking_1994}
(Stimulated Emission Depletion), PALM\cite{betzig_imaging_2006}(Photo-Activated Localisation Microscopy), or STORM\cite{rust_stochastic_2006} (Stochastic Optical Reconstruction
Microscopy). These techniques typically involve additional sophisticated hardware setups and specific sample
preparation protocols, while deep learning based super-resolution can be applied to already existing images, given previous training.

Deep learning based super-resolution has been used in a wide range of applications
so far, for example in satellite imaging, where super-resolution
techniques are employed to enhance the quality of multi-spectral images \cite{collins_deep_2017},
positron emission tomography, where super-resolution methods have been used to reduce the
dosage of radiation\cite{xu_200x_2017,chen_low-dose_2017}, achieving a reduction factor as high as $200$ \cite{xu_200x_2017}, or magnetic resonance imaging\cite{chen_brain_2018}. 
In materials science, Liu {\em et al.} applied this technique to atomic force microscopy to increase image resolution and thereby accelerate imaging \cite{liu_general_2019}. De Haan {\em et al.} \cite{de_haan_resolution_2019} used generative adversarial
networks to improve the resolution of electron microscope images. 
Their emphasis was on mitigating material degradation,
highlighting the importance of super-resolution techniques in preserving sample integrity.

Here, we successfully trained and applied a novel reference-based super-resolution algorithm based on Texture Transformers \cite{yang_learning_2020}, enabling the fast imaging of large areas in the scanning electron microscope. We used two materials, a commercial dual-phase steel, and a 16MnCrS5 case-hardening steel. Besides their industrial relevance both materials were chosen, due to their different and distinct microstructural features.

\section{Methods}

\subsection{Materials}
\label{sec:Materials}

Both dual-phase steels and case hardening steels are pivotal in numerous industrial applications due to their distinct properties and widespread usage in sectors such as automotive manufacturing and heavy machinery, but are quite different when observed in the scanning electron microscope.

\begin{figure}
    \centering
    \includegraphics[width=\textwidth]{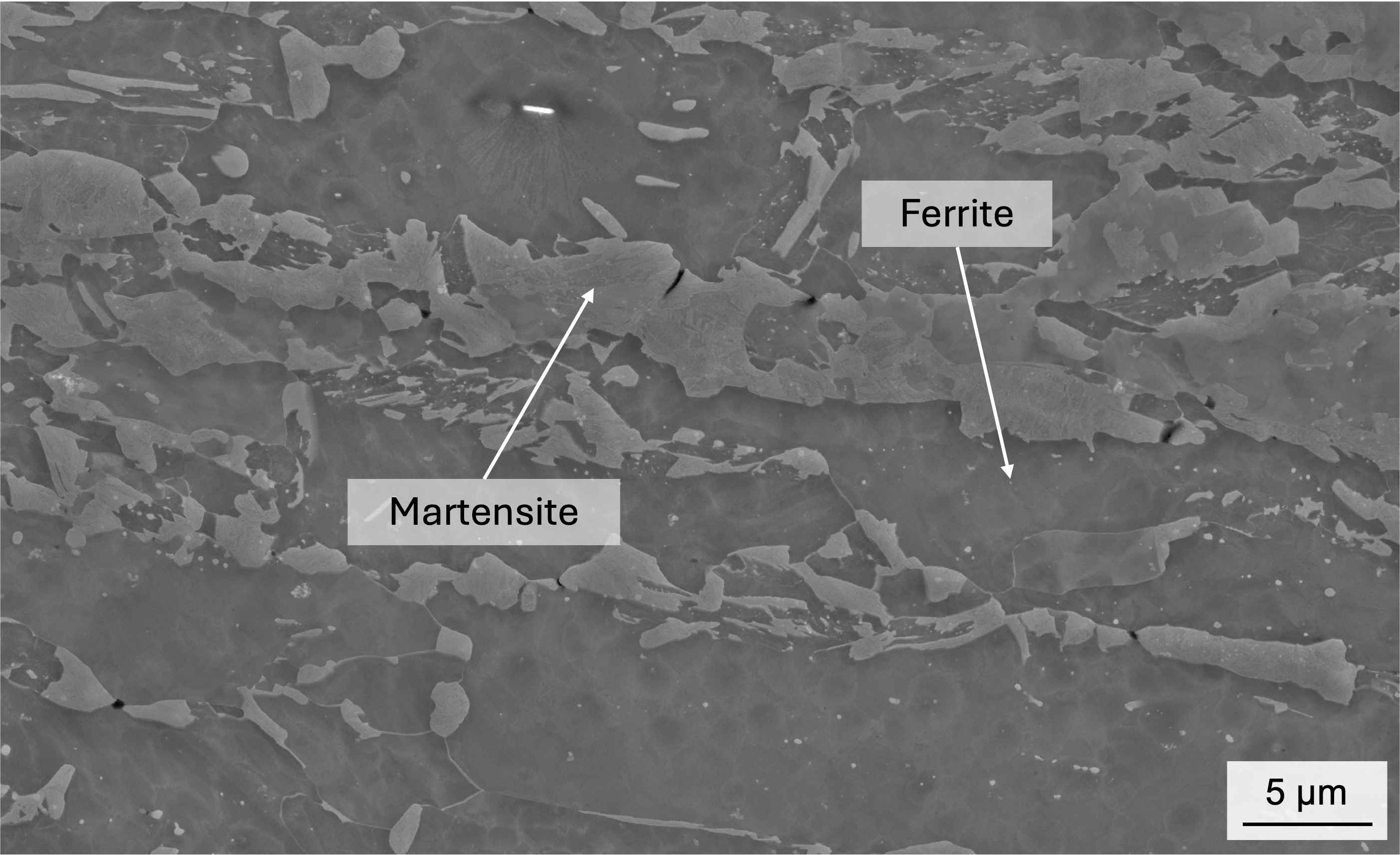}
    \caption{Electron microscope image of a commercial dual-phase steel (DP800), showing the brittle martensite embedded in a softer ferrite matrix.}
    \label{fig:dplabelled}
\end{figure}

\textbf{Dual-Phase Steel}: This material is characterised by its distinctive micro-structure, combining a strong martensite phase within a softer ferrite phase, as shown in \autoref{fig:dplabelled}.
This combination leads to the unique properties of dual-phase steel, such as
its high tensile strength and formability. Together with its low production cost, these properties
make dual-phase steel a favoured material for many applications requiring high fatigue resistance and impact strength. Understanding the mechanisms of damage initiation and propagation during forming processes and in application is crucial for improving material designs.
In this work we use a commercial dual-phase steel of DP800 grade (ThyssenKrupp Steel Europe AG and ArcelorMittal SA, Luxembourg).

\begin{figure}
    \centering
    \includegraphics[width=\textwidth]{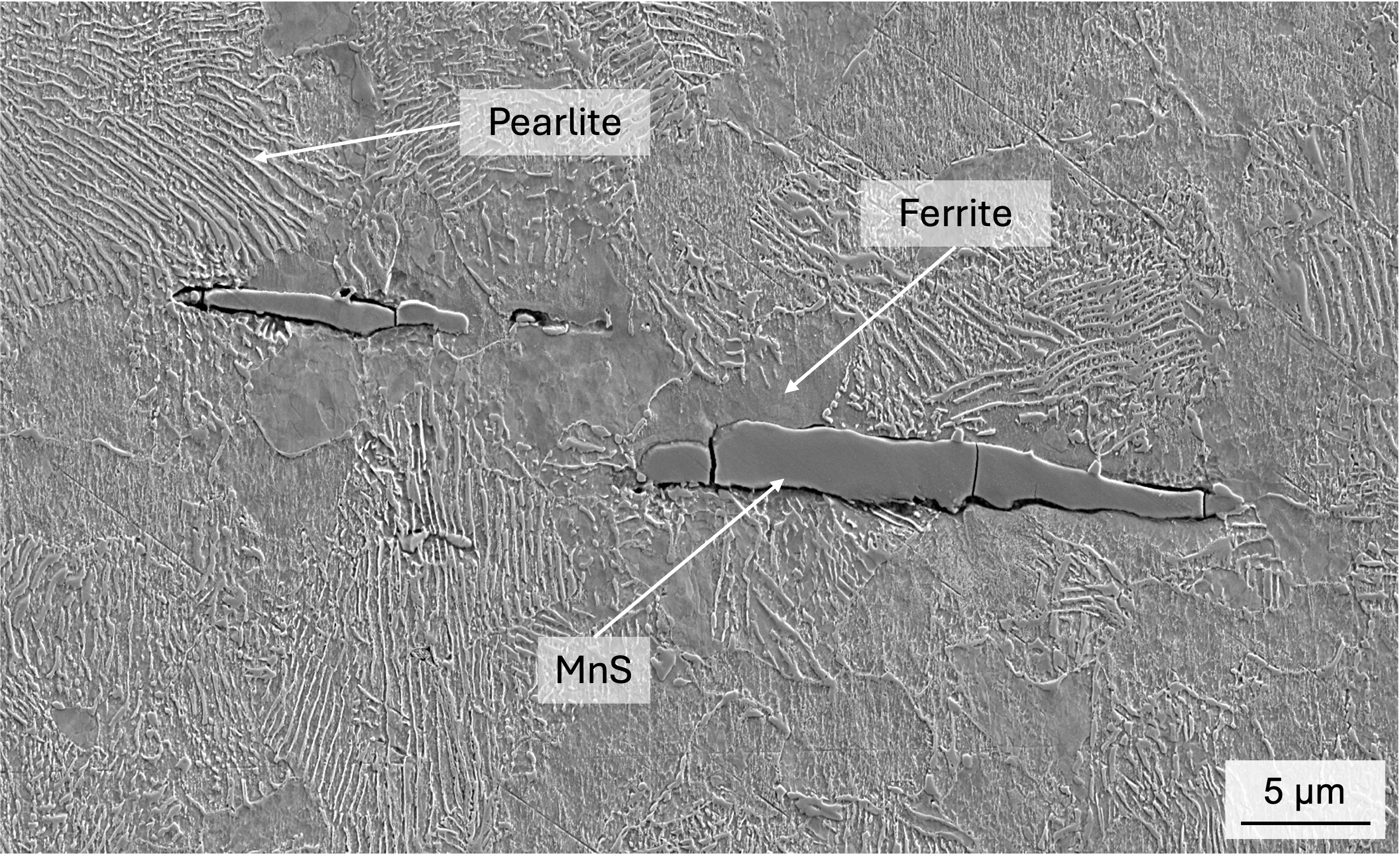}
    \caption{Micro-structure of 16MnCrS5 steel, showing the pearlite lamellae, the softer
    ferrite matrix, as well as larger MnS inclusions.}
    \label{fig:mncrslabelled}
\end{figure}

\textbf{Case-Hardening Steel}: The micro-structure
of 16MnCrS5 steel consists of a pearlitic and
a ferritic phase, as well as dispersed manganese sulfide (MnS) inclusions, as shown in 
\autoref{fig:mncrslabelled}.
Pearlite is characterised by its zebra-striped pattern of alternating ferrite and cementite lamellae. This kind of steel is used primarily for components that require a hard surface to resist wear while maintaining a tough interior to withstand impact loads.
The 16MnCrS5 steel sample is also a commercial steel (Georgsmarienhütte Holding GmbH, Germany).


\subsection{Sample Preparation}
The original steel samples were first cut into a bending sample (dual-phase steel) or a tensile 
test sample (16MnCrS5 steel) using an electrical discharge machine.

To allow imaging of the microstructure of the surface, both samples underwent a similar metallographic preparation process. In a first step, the surface of either sample was ground using sandpaper with grits ranging from 800 to 4000, using water as a cooling agent. Subsequently, the sample surface was mechanically polished using \SI{6}{\micro\meter}, \SI{3}{\micro\meter} and \SI{1}{\micro\meter} water-based diamond suspensions. In this step, a DAC cloth was used for the dual-phase steel sample, while an Alpha cloth was used for the 16MnCrS5  steel sample. Both samples were finished by polishing with a \SI{0.25}{\micro\meter} Oxide Polishing Suspension (OPS) for \SI{1}{\minute}. In a last step, the samples were etched in Nital solution to achieve a topographical contrast, allowing to discriminate between the different phases of the samples under the scanning electron microscope. The dual-phase steel was etched in $1 \%$ Nital solution for \SI{5}{\second}, while the case-hardening steel was etched in $5\%$ Nital solution for \SI{5}{\second}.  

Both samples were deformed in the same microMECHA Proxima testing stage using the appropriate module for the respective sample geometry. The dual-phase steel bending sample was deformed using the three-point bending test module up to a plastic strain of 0.3 at the outer edge of the sample. This deformation step took place before metallographic preparation of the sample. In contrast, the tensile test of the case-hardening steel sample was conducted after metallographic preparation, and terminated at a plastic strain of $7\%$. After preparation and deformation the samples were mounted in the scanning electron microscope chamber for the analysis of the prepared surface. 

\subsection{Image Acquisition}
\label{sec:imageAcquisition}

The electron micrographs were acquired using a TESCAN CLARA (Tescan Group, Czech Republic) scanning electron microscope 
with secondary electrons for detection. For the dual-phase steel sample, an accelerating voltage of \SI{20}{\kilo\volt} and a beam current of \SI{3}{\nano\ampere} was used. For the 16MnCrS5 sample, an accelerating voltage of \SI{10}{\kilo\volt} and a beam current of \SI{1}{\nano\ampere} was used. 

\begin{figure}
    \centering
    \includegraphics[width=\textwidth]{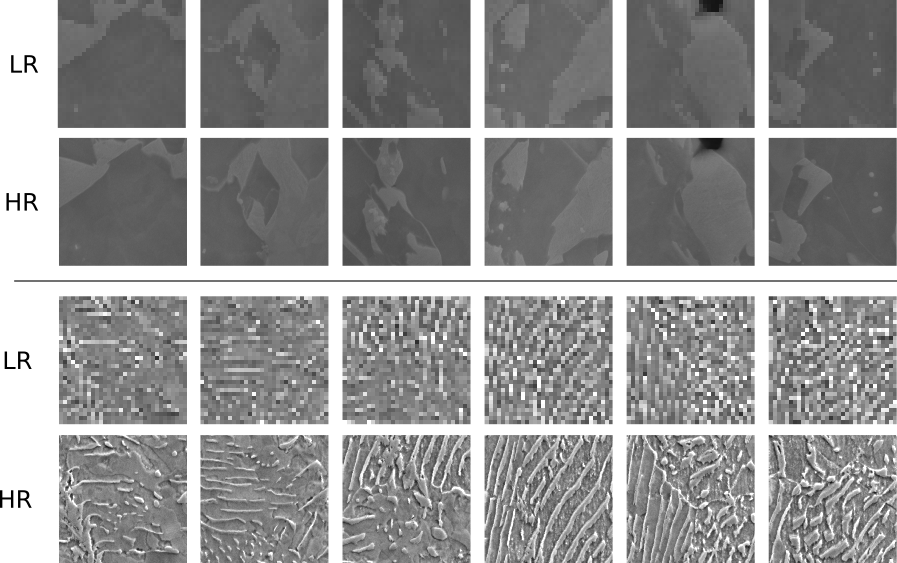}
    \caption{Examples for pairs of high-resolution (HR) and low-resolution (LR) images used
    as training data. top: Dual-phase steel, bottom: 16MnCrS5 steel}
    \label{fig:image_pairs}
\end{figure}

The training of the super-resolution algorithm
requires two sets of images, each of which showing the same region of the experimental sample:
One set of images is recorded at the target resolution required for a later analysis. For the dual-phase steel dataset we chose this resolution to be the same 
as in \cite{kusche_large-area_2019}.
For the 16MnCrS5  steel, we chose 
a resolution of \SI{32.5}{\micro\meter} per pixel as in \cite{wollenweber_automated_2023}. A second set of images is recorded at a lower resolution,
this is the resolution that we aim to use in the later acquisition of the actual images. These images still need to contain enough information for the subsequent resolution enhancement to work at this resolution. Intuitively, the larger the difference between the high-resolution and the low-resolution, the higher the chance that the algorithm
will infer image features that are not present in the physical sample, as many low-resolution images can correspond to the same high-resolution
image, and the larger the difference in resolution, the more potential low-resolution
images can correspond to the same high-resolution image.
On the other hand, in order to achieve a substantial improvement in image acquisition
speed, we need to choose \textit{some} difference between the low- and high-resolution images. Therefore, in this work,  we chose a factor of \SI{4}{} in the resolution difference for both materials, see \autoref{fig:image_pairs} for exemplary images. This resolution difference in both dimensions would then bring the  acquisition time down by a factor of \SI{16}{}.

During dataset generation, one factor that can strongly influence the dataset quality is the window size in pixel that the microscope records the images in. As effects such as drift accumulate over time, the larger the window size, the larger the potential misalignment between the low-resolution and high-resolution images in subsequent scans. The lower end of the window size, that is still useful, is the input size used in the neural network. 
We, therefore, chose for the initial training of our resolution enhancement network a window size of $128\times 128$ pixel, corresponding to the perceptive window size of the used network, for the dual-phase steel sample and recorded $25,000$ image pairs. Additionally, $256$ reference images of $300\times 300$ pixel at
the target resolution were recorded. These reference images
play a pivotal role in the details of the super-resolution algorithm explained later in \autoref{sec:srAlgorithm}. On the other end is the largest window size that the microscope is capable of recording in a single session. 
We test the other extreme by recording with an window size of $4096\times 4096$ pixel, and obtain $1$ image on the dual-phase steel sample for testing the capability of the network to predict larger micrographs, and $160$ images on the 16MnCrS5 sample to train and evaluate on a second microstructure.
From the 16MnCrS5 micrographs, we select one micrograph to be enhanced in full for testing, three are selected to be used as reference images and we divide the remaining images into patches of size $128\times 128$ pixel and $300\times 300$ pixel. These patches are then realigned by searching for the position in the high-resolution image, that after down sampling to the low-resolution maximises the SSIM score, explained in \autoref{sec:metrics}. An overview of the recorded datasets can be found in \autoref{tab:datasets}.

\begin{table}[ht]
\centering
\caption{Overview of recorded datasets. The DP-large dataset was only recorded for inference and has therefore no training and validation data.}
\label{tab:datasets}
\begin{tabular}{lccc}
\toprule
Name & DP-small & DP-large & 16MnCrS5 \\ 
\midrule
Recorded image pairs &  \SI{25,000}{} & \SI{1}{} & \SI{160}{} \\ 
Low-resolution window size &  \SI{32}{} $\times$ \SI{32}{} & \SI{1024}{} $\times$ \SI{1024}{} & \SI{1024}{} $\times$ \SI{1024}{} \\ 
High-resolution window size &  \SI{128}{} $\times$ \SI{128}{} & \SI{4096}{} $\times$ \SI{4096}{} & \SI{4096}{} $\times$ \SI{4096}{} \\ 
\midrule
Recorded reference images &  \SI{256}{} & - & - \\ 
Window size &  \SI{300}{} $\times$ \SI{300}{} & - & - \\
\midrule
Number of training pairs &  \SI{20,000}{} & - & \SI{1,600}{} \\ 
Number of validation pairs &  \SI{2,500}{} & - & \SI{200}{} \\ 
Number of test pairs &  \SI{2,500}{} & \SI{1,700}{} & \SI{200}{} \\ 
Number of reference images & \SI{256}{} & \SI{256}{} (Same as DP-small) & \SI{256}{} \\
\bottomrule
\end{tabular}
\end{table}

\subsection{Image Preprocessing}
Deep learning applications in computer vision are notorious for requiring large datasets. For example, many neural networks for imaging
tasks are trained using the ImageNet dataset \cite{deng_imagenet_2009} that consists of more than 14 million images.
While such datasets are publicly available, the images mainly show every-day objects 
and are, therefore, not directly suitable for scientific applications.
However, creating a similarly large dataset for each specific application
domain is not possible due to the enormous amount of effort that would be needed
to create these images and, potentially, add labels with their content.
However, this can be mitigated by using transfer learning \cite{west_theoretical_2007}, by using a neural network trained on a different task and then only use a smaller dataset to train it for the desired application. Here we use the trained weights on the CUFED5 dataset \cite{zhang_image_2019} available from the original texture transformer for super resolution (TTSR) publication \cite{yang_learning_2020}.

However, the original TTSR network was trained on colour images, consisting of three 
colour coordinates ${R,G,B}$ for each spatial coordinate, while the scanning electron microscope only records an intensity, i.e. only one channel. Therefore, as a first step we convert the grey scale images to the expected colour format. Then additionally we re-scale the colour information to the interval $[-1,1]$.

\subsection{Super-Resolution using Texture Transformers}
\label{sec:srAlgorithm}
\subsubsection{Super-Resolution Algorithm}

Super-resolution methods based on machine learning generally fall in one of two categories: Either only a single input image is algorithmically enhanced, which is commonly referred to as single-image super-resolution \cite{dong_image_2015, dong_accelerating_2016, ledig_photo-realistic_2017,lu_transformer_2022}.
Alternatively, in reference-based super-resoltuion a set of reference images is used in addition to 
the individual input image \cite{yang_learning_2020, cao_reference-based_2022, zhang_image_2019}. Our approach
falls into the latter category. These reference images provide additional information that can help in reconstructing high-resolution information.

\begin{figure}[h]
    \includegraphics[width=\textwidth]{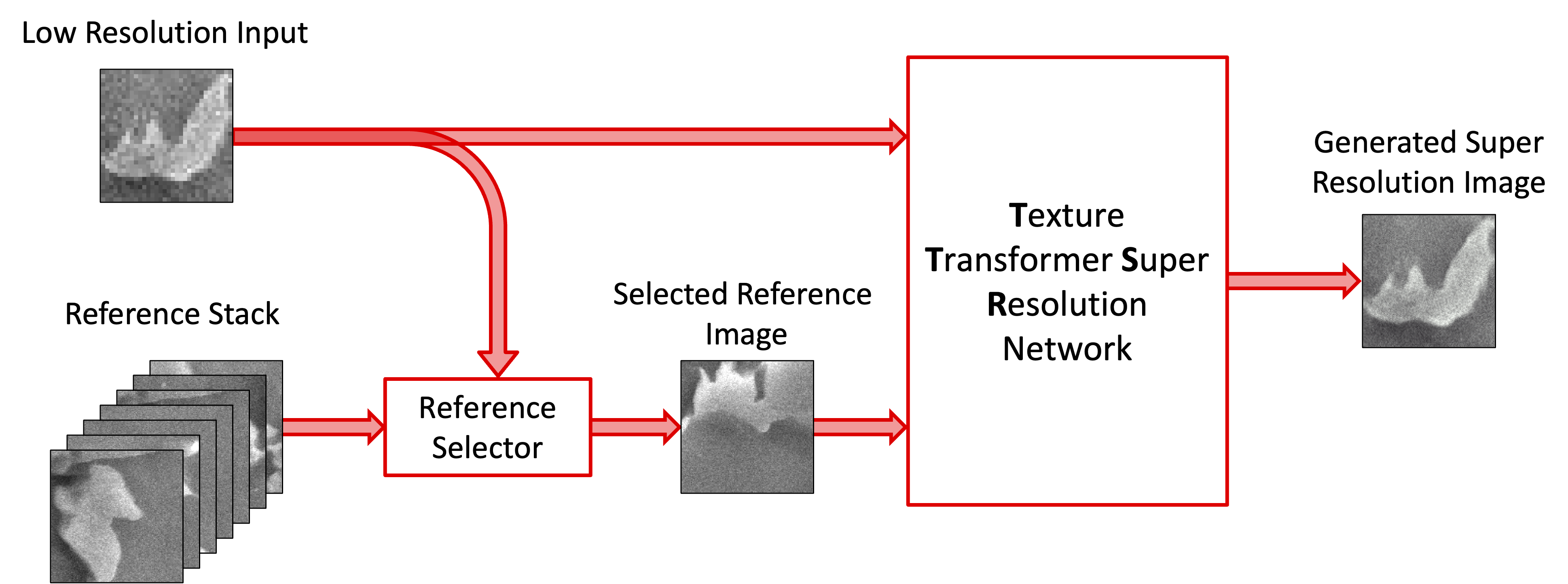}
    \caption{Modified Texture Transformer Model}
    \label{fig:modifiedTextureTransformer}
\end{figure}

\autoref{fig:modifiedTextureTransformer} illustrates the overall approach of the TTSR algorithm during inference.
In a first step, we need to select a suitable reference image from the stack
of reference images. 
For this purpose, the low-resolution image is analysed in the
reference selector: Using bi-cubic interpolation, the low-resolution image is 
up-sampled to high-resolution. In order to be able to compare the similarity between the high-resolution reference images and the low-resolution input, high-resolution information from the reference image needs to be removed. This is done by down-sampling the reference images to the low-resolution and then up-sampling them again using bi-cubic interpolation. Both the input image and the transformed reference image are then passed to a learnable texture extractor, a component of the network, and the output of an intermediate layer is used as a representation of each image. The similarity is then determined by calculating the cosine similarity between these vectors, which is defined as 
\begin{equation}
    \cos \alpha = \frac{\vec{v}\cdot \vec{w}}{\|\vec{v}\|\|\vec{w}\|}
\end{equation}


The reference image with the highest similarity to the input is then chosen for further processing. Then, in a second step, both the low-resolution input image, as well as the selected reference image are used as input to the super-resolution network. This network is based on the texture transformer\cite{yang_learning_2020}.

For the training process we select the reference images with the highest similarities ahead of time in order to speed up training, as the process of recalculating and storing the vector representations of all the reference images can be computationally intensive.

\subsubsection{Model Training}

We divide each dataset in to $80\%$ training data, $10\%$ validation data, and $10\%$ test data. Corresponding to $20,000$, $2,500$, and $2,500$ high-resolution, low-resolution, reference image triplets respectively for the dual-phase steel dataset and $1,600$, $200$, $200$ high-resolution, low-resolution, reference image triplets for the 16MnCrS5 dataset.


One of the pivotal components in any supervised machine learning model is the loss function,
as its definition determines what the algorithm should learn.
Na\"ively, one might assume that the pixel-wise difference between the super-resolution and the corresponding high-resolution, may suffice, commonly referred to as the reconstruction loss and defined as
\begin{equation}
    \mathcal{L}_{rec} = \frac{1}{CHW}\|I^{HR}-I^{SR}\|_1
\end{equation}
Here, C, H, and W are the number of channels, the height, and the width, $I$ represents the image, and $\|\ldots\|_1$ is the $L_1$ norm of the difference.
However, while this difference plays a key role in asserting the resulting image quality, this metric alone is not sufficient, as the reconstruction loss inherently prioritises minimising pixel-wise differences often at the expense of high-frequency details, networks trained with only this loss function tend to produce overly smooth images. Therefore, as in \cite{yang_learning_2020}, we additionally use two further loss components, perceptual loss\cite{johnson_perceptual_2016} and adversarial loss\cite{goodfellow_generative_2014}.
The perceptual loss calculates the difference in an intermediate layer of a trained neural network between the generated super-resolution image and the ground truth \cite{johnson_perceptual_2016} and has been specifically developed for the application of neural networks in super-resolution algorithms
\begin{equation}
    \mathcal{L}_{per} = \frac{1}{C_iH_iW_i}\|\phi_i^{vgg}(I^{SR})-\phi_i^{vgg}(I^{HR})\|_2^2 +
    \sum_j\frac{1}{C_jH_jW_j}\|\phi_j^{lte}(I^{SR})-T\|_2^2
\end{equation}
$\phi_i^{vgg}$ is the VGG19 network deprecated after the $i$th layer, $C_i$, $H_i$, and $W_i$ are the number of channels, the height and width of that layer, correspondingly $\phi_j^{lte}$ is the learnable texture extractor deprecated after the $j$th layer, $C_j$, $H_j$, and $W_j$ are the number of channels, the height and width of the $j$th layer. Here we choose $i$ to be the layer after the $13$th convolutional layer and sum over $j$ to be chosen after the first convolutional layer, the third convolutional layer, and after the fifth convolutional layer, as done in \cite{yang_learning_2020}. 
Instead of the $L_1$ norm, this component uses the $L_2$ norm.
This contribution to the loss function ensures that high-level features are similar between the 
image after the super-resolution algorithm has been applied and ground truth images.

Finally, the adversarial loss is based on the idea behind generative adversarial networks (GANs)\cite{goodfellow_generative_2014}. Adversarial networks typically consist of two competing networks, where
the generator network creates new images and the discriminator network tries to 
tell the output from the generator apart from the ``real'' images in the training data (ground truth).
As in \cite{yang_learning_2020} we chose the Wasserstein loss \cite{arjovsky_wasserstein_2017} given by
\begin{equation}
    \mathcal{L}_{adv} = \mathop{\mathbb{E}}_{\tilde{x}\sim \mathbb{P}_g}[D(\tilde{x})] -
    \mathop{\mathbb{E}}_{x\sim \mathbb{P}_d}[D(\tilde{x})]
    + \lambda \mathop{\mathbb{E}}_{\hat{x}\sim \mathbb{P}_{\hat{x}}}[(\|\nabla_{\hat{x}}D(\tilde{\hat{x}})\|_2-1)^2]
\end{equation}
where $D$ is the discriminator, $\mathbb{E}$ is the expectation value over $\mathbb{P}_g$, the distribution over generated super-resolution images, $\mathbb{P}_d$ the distribution of ground truth images, $\mathbb{P}_{\hat{x}}$, the distribution of all possible images, $\lambda$ is a parameter, determining the degree of gradient penalisation, here set to $10$.

The final loss function is a weighted combination of the three contributions detailed above
\begin{equation}
    \mathcal{L}_{tot} = \lambda_{rec} \mathcal{L}_{rec} + \lambda_{per} \mathcal{L}_{per} + \lambda_{adv} \mathcal{L}_{adv}
    \label{eq:loss_function}
\end{equation}
The individual weights $\lambda_i$ need to be tuned as part of the network training
to balance the strength with which they enter the training of the network. The contribution
of the various components also depends on the overall progress of the training.
Using the complete loss function from the beginning of the training can lead to unstable behaviour, where the competing contributions
can lead to undesired outputs. This is well known from generative adversarial networks \cite{metz_unrolled_2017}. To mitigate this behaviour, we follow the approach 
by Yang et. al \cite{yang_learning_2020} and first train the network only with the 
reconstruction loss and then add the other contributions later. This has the additional
benefit, that---as we use networks pre-trained on real world colour images---these networks
have time to adapt to gray-scale scanning electron micrographs.

\begin{figure}
    \centering
    \includegraphics[width=\textwidth]{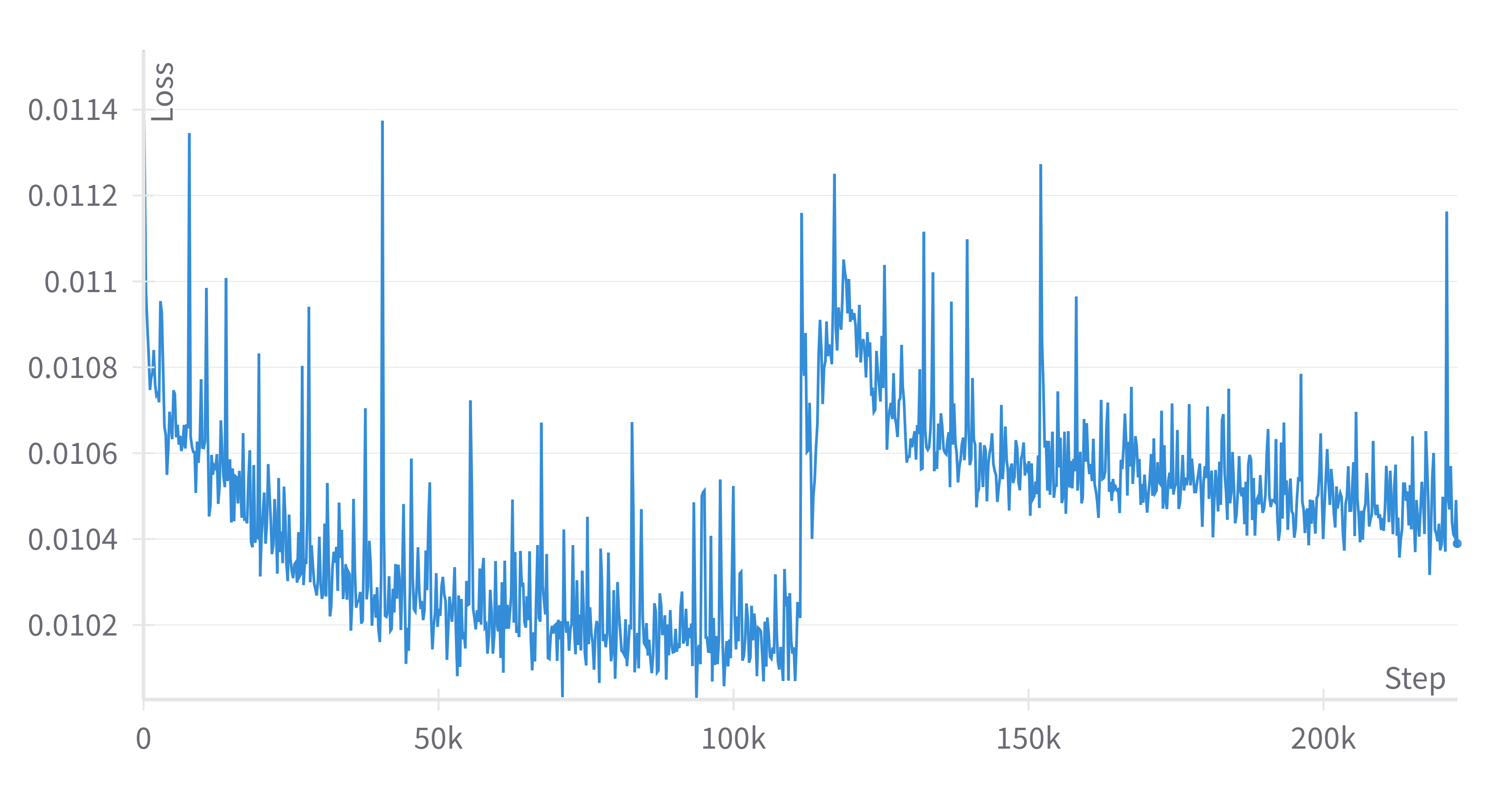}
    \caption{Loss function during training progression. It quickly converges to a value at around $0.0102$. After step $100,000$ the remaining loss functions are added to the reconstruction loss, resulting in a slight increase.}
    \label{fig:loss}
\end{figure}

The further technical details of the training are as follows:
We used a cyclic learning rate scheduler \cite{smith_cyclical_2017} and Adam optimiser \cite{kingma_adam_2017}.
The training process was executed on an NVIDIA Tesla V100 SXM2 16 GB on the RWTH ITC high-performance cluster, which took \SI{20}{\hour} for the dual-phase steel dataset and \SI{3}{\hour} for the 16MnCrS5 dataset. The resulting
loss function during training is illustrated in \autoref{fig:loss} where the different
phases of the training regime can be clearly seen.

\subsection{Evaluation Metrics}
\label{sec:metrics}
\noindent Evaluating the performance of our super-resolution algorithm necessitates
the use of reliable metrics that can provide
insights into the quality of the reconstructed images. To this end we employ two common metrics used
in deep learning super-resolution.\\

\begin{enumerate}
    \item \textbf{Structural Similarity Index Measure (SSIM)}:
          The SSIM quantifies the
          structural integrity preserved in the images resulting from the machine-learning algorithm compared to the (high-resolution) ground-truth data. This metric compares the
          luminance, contrast, and structure between the reconstructed and high-resolution images. It takes values between \SI{-1}{} and \SI{1}{}, where
          \SI{1}{} would be a perfect match and \SI{-1}{} a complete mismatch \cite{wang_image_2004}.
    \item \textbf{Peak Signal-to-Noise Ratio (PSNR)}: The PSNR is, essentially, an extension of the mean squared error to images and measures the ratio between the maximum possible power of
          the signal, here the maximum pixel value, and the corrupting noise, the squared difference between the pixel values of the predicted image and the high-resolution ground truth.
          The value of PSNR is expressed in \SI{}{dB}, the higher the value the higher
          the similarity between the images.
\end{enumerate}

\subsection{Interpolation}

An alternative to deep learning methods are interpolation based methods to increase the size of an image. These interpolation algorithms assume a function underlying the pixel values, determine the parameters of these functions and then evaluate at the missing pixel positions. Nearest neighbour interpolation assumes a step function, missing pixel are then the pixel value of the nearest known pixel value. Bi-cubic interpolation assumes cubic functions in both dimensions, while Lanczos interpolation \cite{duchon_lanczos_1979} uses $\mathrm{sinc(x)} = \frac{\sin(x)}{x}$ functions. While these algorithms can increase the image size, they have no additional information. Here they serve as a baseline to compare the trained networks against.




\section{Results}
In order to be able to confidently apply resolution enhancement with our trained texture transformer networks, we need to establish, that they substantially improve over conventional interpolation methods. Therefore we first present interpolation results and thereby establish a baseline for comparison. Then we apply the TTSR network to the dual-phase steel dataset, which was recorded with a small window size. The trained network is then applied to a larger dual-phase steel micrograph, without any additional training of the network. Lastly, we apply the super-resolution network to our second material, 16MnCrS5. We compare both in terms of image similarity metrics, here SSIM and PSNR, and also provide a visual comparison between input, ground-truth, interpolation and output.

\subsection{Baseline}
\begin{figure}[h!]
    \centering
    \includegraphics{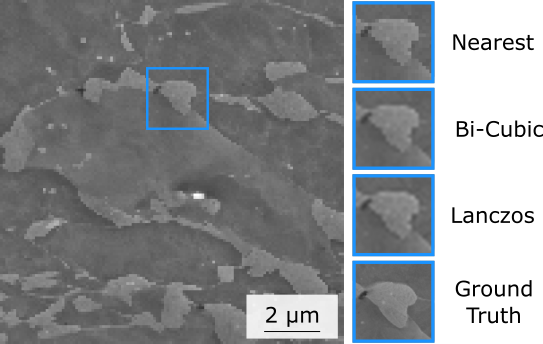}
    \caption{Comparison of nearest neighbour, bi-cubic, and Lanczos interpolation with the high-resolution ground truth on a dual-phase steel scanning electron micrograph.}
    \label{fig:interpolationComparison}
\end{figure}

\autoref{fig:interpolationComparison} shows a visual comparison
between nearest, bi-cubic and Lanczos interpolation. Nearest interpolation leads to block like structures in the input, while bi-cubic and Lanczos interpolation both produce blurry images. Bi-cubic and Lanczos interpolation produce visually very similar images. \autoref{tab:interpolation_performance_metrics} displays the similarity metrics for bi-cubic and Lanczos interpolation on the dual-phase steel dataset. While both interpolation methods achieve similar similarity metrics, bi-cubic interpolation performs better on both dual-phase steel datasets.

\begin{table}[ht]
\centering
\caption{Similarity metrics for bi-cubic and Lanczos interpolation on the dual-phase steel data.}
\label{tab:interpolation_performance_metrics}
\begin{tabular}{lcc}
\toprule
Metric  & Bi-Cubic & Lanczos \\ \midrule
SSIM (DP-small) &\SI{0.622}{} $\pm$ \SI{0.002}{}& \SI{0.574}{} $\pm$ \SI{0.002}{}                \\ 
PSNR (DP-small) &\SI{25.57}{\decibel} $\pm$ \SI{0.05}{\decibel}& \SI{24.38}{\decibel} $\pm$ \SI{0.04}{\decibel}                \\ 
\midrule
SSIM (DP-large) &\SI{0.559}{} $\pm$ \SI{0.005}{}& \SI{0.546}{} $\pm$ \SI{0.005}{}                \\ 
PSNR (DP-large) &\SI{22.1}{\decibel} $\pm$ \SI{0.1}{\decibel}& \SI{21.8}{\decibel} $\pm$ \SI{0.1}{\decibel}                \\
\end{tabular}
\end{table}

\subsection{Evaluation of Network Performance on Dual-Phase Steel Dataset}
\begin{figure}
    \centering
    \includegraphics[width=0.5\textwidth]{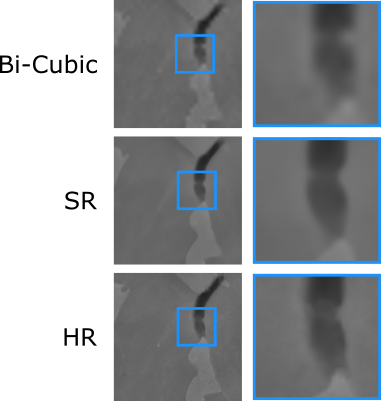}
    \caption{Exemplary boundary decohesion from the test set of the dual-phase dataset, displayed after bi-cubic interpolation, super-resolution and high-resolution. The larger sections look very similar in all three cases. Additionally displayed is a smaller region zoomed in, to emphasise the difference between interpolation and super-resolution.}
    \label{fig:dpBoundaryInterpolationComparison}
\end{figure}

We start with the dataset consisting of images recorded directly in the input size of the network as described in \autoref{sec:imageAcquisition}. \autoref{tab:dpSmallPerformanceMetrics} depicts the similarity metrics for bi-cubic interpolation, Lanczos interpolation and super-resolution. Contrary to the expectations, the bi-cubic interpolation performs better than the Lanczos interpolation and comes close to the trained network for super-resolution. \autoref{fig:dpBoundaryInterpolationComparison} provides an example of a damage site, a boundary decohesion, after bi-cubic interpolation, super-resolution and in the high-resolution ground truth. While the entire patch looks similar in all three cases, the zoomed section shows significant differences. The interpolated image appears blurred and the edges of the damage site are smeared. The super-resolution image, on the other hand, shares a  higher degree of similarity with the high-resolution image. Features such as the edge of the damaged area and the small martensite island at the bottom of the zoomed-in section are sharper than in the interpolated image.

\begin{figure}
    \centering
    \includegraphics[width=0.7\textwidth]{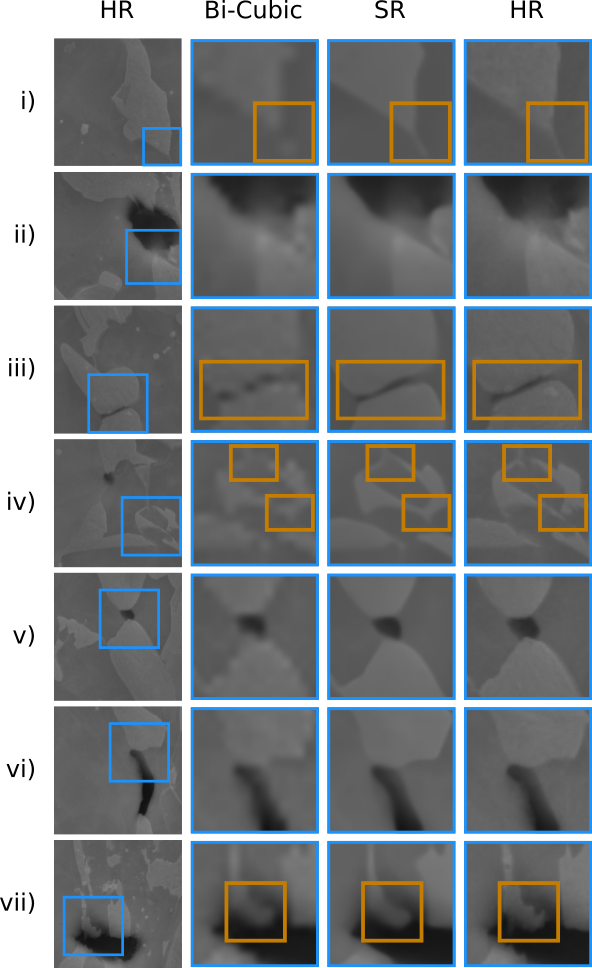}
    \caption{Scanning electron micrographs of features in the dual-phase steel microstructure. Each row corresponds to a different feature. i) Martensite islands, ii) inclusions, iii) martensite cracks, iv) interface decohesions, v) notch effect, vi) boundary decohesion, and vii) evolved damage sites.}
    \label{fig:largerMicrostructurePlot}
\end{figure}

\autoref{fig:largerMicrostructurePlot} shows a variety of patches from the test data set. Each line contains a patch in high-resolution and then a zoomed region after bi-cubic interpolation, super-resolution and in high-resolution. 
Since a large body of research deals with damage in dual-phase steels \cite{tian_understanding_2024, wollenweber_damage_2023, medghalchi_damage_2020, kusche_large-area_2019}, patches showing different damage sites and inclusions were selected in addition to a martensite island. 
Further examples for each category can be found in \autoref{sec:moreExamples}. Overall, the bi-cubic interpolated areas appear blurred compared to the super-resolution results and the high-resolution ground truth. A martensite island is shown in the i). Not only the clear interface to the ferrite is correctly reconstructed, but also the small bridge attached to the lower right corner of the island is correctly shown in the super-resolution image. ii) depicts an inclusion site. The lower section of the void is emphasised here, where the super-resolution image again displays distinct borders of the different phases, close to a larger void.  In iii), we observe a martensite crack. In the interpolated image, the crack appears discontinuous, with possible small bridges between the martensite sections. The super-resolution correctly reconstructs the crack as one continuous void. An example of the problem caused by this network is visible in the subsequent row, iv). While it depicts interface decohesion, we emphasise the smaller martensite structure in the lower right part. Although the reconstructed structure looks similar to the high-resolution structure, differences exist. The shape of the martensite in the lower right deviates from the ground truth, and the section above appears as a continuous piece, while small openings are present in the high-resolution ground truth. In v) and vi), a notch effect site and a boundary decohesion site are shown. In both cases, the super-resolution network generates different phases compared to the interpolation. Finally, an evolved damage site is presented in vii). Similar to the interface decohesion example, fine structures exist that the network is not capable of reconstructing.

\begin{table}[ht]
\centering
\caption{Similarity metrics for bi-cubic, Lanczos interpolation, and the trained super-resolution network on the DP-small dataset.}
\label{tab:dpSmallPerformanceMetrics}
\begin{tabular}{lccc}
\toprule
Metric &  Bi-Cubic&Lanczos & TTSR \\ \midrule
SSIM &  \SI{0.622}{} $\pm$ \SI{0.002}{}&\SI{0.574}{} $\pm$ \SI{0.002}{}                & \SI{0.626}{} $\pm$ \SI{0.002}{} \\ 
PSNR &  \SI{25.57}{\decibel} $\pm$ \SI{0.05}{\decibel}&\SI{24.38}{\decibel} $\pm$ \SI{0.04}{\decibel}                & \SI{25.96}{\decibel} $\pm$ \SI{0.04}{\decibel} \\ 
\bottomrule
\end{tabular}
\end{table}

\subsection{Application to Larger Dual-Phase Steel Micrographs}
\begin{figure}
    \centering
    \includegraphics[width=\textwidth]{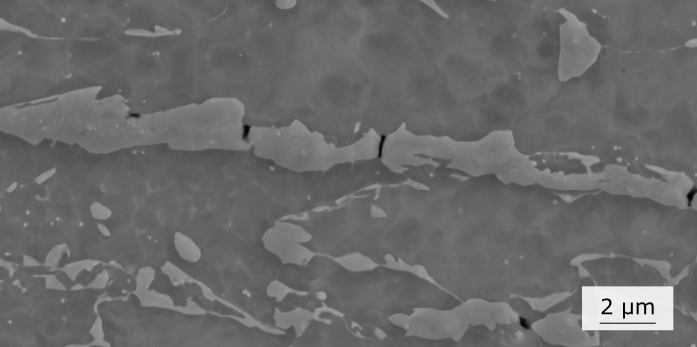}
    \caption{Example of super-resolution patches, stitched together to larger micrograph.}
    \label{fig:stitchedSRMicrograph}
\end{figure}

\begin{figure}
    \centering
    \includegraphics[width=0.68\textwidth]{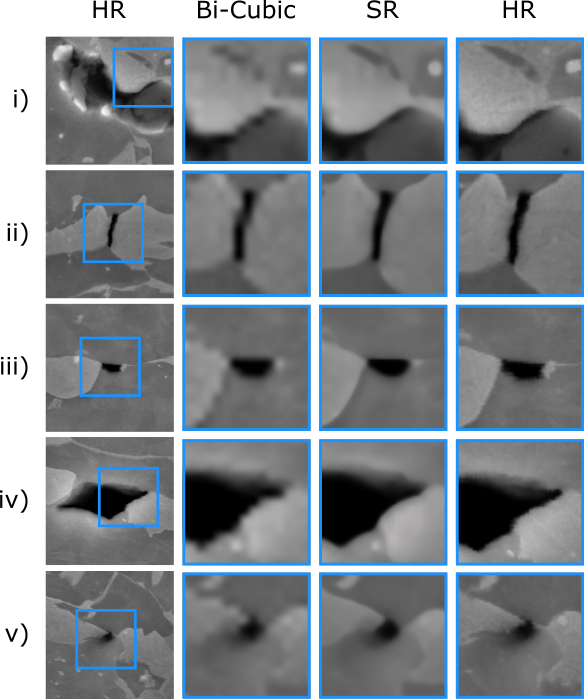}
    \caption{Inclusions and damage sites, together with a zoomed window for each example, taken from the larger dual-phase steel micrographs, in bi-cubic interpolation, super-resolution and high-resolution. i) inclusion, ii) martensite crack, iii) boundary decohesion, iv) evolved damage site, v) notch effect.}
    \label{fig:dpsrdamage}
\end{figure}

To evaluate the ability of the trained network to enhance larger micrographs, we test its predictions of a micrograph recorded with a window size of $4096\times 4096$ pixels (DP-large) and compare it to both bi-cubic interpolation and the high-resolution ground truth. Due to the larger window size, we need to divide the larger micrograph into smaller patches, processable by the network. After processing, the networks predictions are assembled together to the size of the original micrograph. As artefacts at the borders of the prediction can occur, only the central part of the prediction is used for this stitching process. A small section of the stitched super-resolution micrograph is depicted in 
\autoref{fig:stitchedSRMicrograph}. This section is $1024\times 512$ pixels large which equals to $8 \times 4$ images of the output size of the network. Through the process described above, no stitching artefacts can be seen. Due to accumulated drift, the low-resolution patches need to be realigned with the high-resolution patches in order to allow the calculation of similarity metrics and direct visual comparison.  In addition, we selected inclusions and damage sites, which are shown for bi-cubic interpolation and super-resolution in \autoref{fig:dpsrdamage}. Again in order to allow for a comparison, we show the high-resolution ground truth. Even though the images here were not taken under the same conditions, as the training dataset, the predictions of the network resemble the high-resolution ground truth. Similar to \autoref{fig:largerMicrostructurePlot}, bi-cubic interpolation produces blurry images, while the networks reconstructions contain sharp edges. This is especially visible in \autoref{fig:dpsrdamage} i), iv) and v).

\begin{table}[ht]
\centering
\caption{Quantitative performance evaluation the DP-large dataset.}
\label{tab:performance_metrics}
\begin{tabular}{lccc}
\toprule
Metric & Bi-Cubic &Lanczos & TTSR \\ 
\midrule
SSIM&  \SI{0.559}{} $\pm$ \SI{0.005}{}&\SI{0.546}{} $\pm$ \SI{0.005}{}                & \SI{0.574}{} $\pm$ \SI{0.005}{} \\ 
PSNR&  \SI{22.1}{\decibel} $\pm$ \SI{0.1}{\decibel}&\SI{21.8}{\decibel} $\pm$ \SI{0.1}{\decibel}                & \SI{22.2}{\decibel} $\pm$ \SI{0.1}{\decibel} \\ 
\bottomrule
\end{tabular}
\end{table}

\subsection{Transferability to 16MnCrS5 Micrographs}
Ideally, the trained network can be directly applied to other applications or materials. To test if and how much adaption to a new material is necessary, we apply the method on 16MnCrS5 as discussed in \autoref{sec:Materials}.

\subsubsection{Baseline}
\begin{figure}
    \centering
    \includegraphics[width=\textwidth]{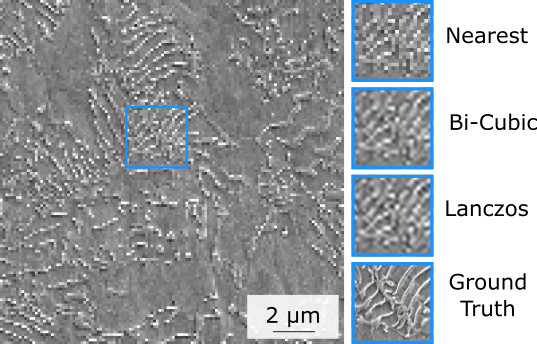}
    \caption{Comparison of nearest neighbour, bi-cubic, and Lanczos interpolation with the high-resolution ground truth on a 16MnCrS5 scanning electron micrograph.}
    \label{fig:16MnCrS5Interpolation}
\end{figure}

Before assessing the networks performance, we again establish a baseline for comparison. In \autoref{fig:16MnCrS5Interpolation} we display nearest, bi-cubic, and Lanczos interpolation alongside the high-resolution ground truth images. The cementite lamellae are not visible in the nearest neighbour interpolation and are barely visible both for bi-cubic and Lanczos interpolation. \autoref{tab:16MnCrS5InterpolationMetrics} presents the similarity metrics for Lanczos and bi-cubic interpolation. Again bi-cubic interpolation performs better than Lanczos interpolation. Therefore we choose it for later comparisons.

\begin{table}[ht]
\centering
\caption{Similarity metrics for bi-cubic and Lanczos interpolation on the aligned 16MnCrS5 dataset.}
\label{tab:16MnCrS5InterpolationMetrics}
\begin{tabular}{lcc}
\toprule
Metric &  Bi-Cubic & Lanczos \\ \midrule
SSIM &  \SI{0.476}{} $\pm$ \SI{0.003}{}&\SI{0.467}{} $\pm$ \SI{0.003}{} \\ 
PSNR &  \SI{18.4}{\decibel} $\pm$ \SI{0.1}{\decibel}&\SI{18.1}{\decibel} $\pm$ \SI{0.1}{\decibel} \\ 
\bottomrule
\end{tabular}
\end{table}
\subsubsection{Direct Application of Trained Network to 16MnCrS5 Micrographs}
\begin{figure}
    \centering
    \includegraphics[width=\textwidth]{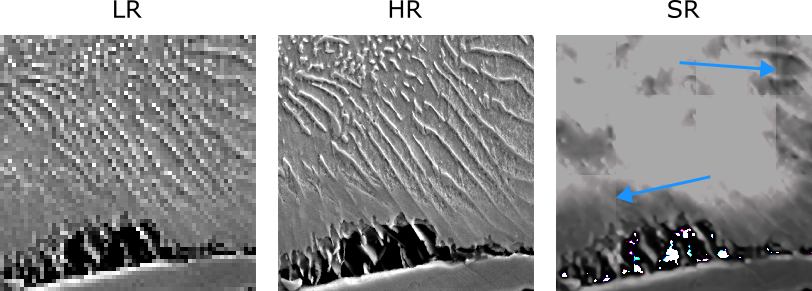}
    \caption{TTSR network trained on dual-phase steel dataset applied to 16MnCrS5 micrograph without finetuning, shown together with the low-resolution input and high-resolution ground truth. Stitching artefacts are indicated with blue arrows.}
    \label{fig:transferNoFinetuning}
\end{figure}

First we test whether the trained network can directly be applied to 16MnCrS5 images, by just supplying 16MnCrS5 reference images. The direct application of the network trained on the dual-phase steel dataset leads to overly bright images, with a mean pixel value of $202$, while the ground truth image has a mean pixel value of $133$. In order check whether this is a simple brightness/contrast issue, we first re-scale the pixel values of the networks prediction, to the high-resolution ground truth. The result, together with the low-resolution input and high-resolution ground truth can be seen in \autoref{fig:transferNoFinetuning}. The predicted image still has over-lit sections with colour artefacts at their borders, and bears little resemblance with the high-resolution ground truth besides rough structures. Furthermore, the features at the borders of predictions do not align leading to stitching artefacts, which can be seen as straight lines both horizontal and vertical direction in the super-resolved image, marked with blue arrows.

\subsubsection{Finetuning with 16MnCrS5 Data}
\begin{figure}
    \centering
    \includegraphics[width=0.68\textwidth]{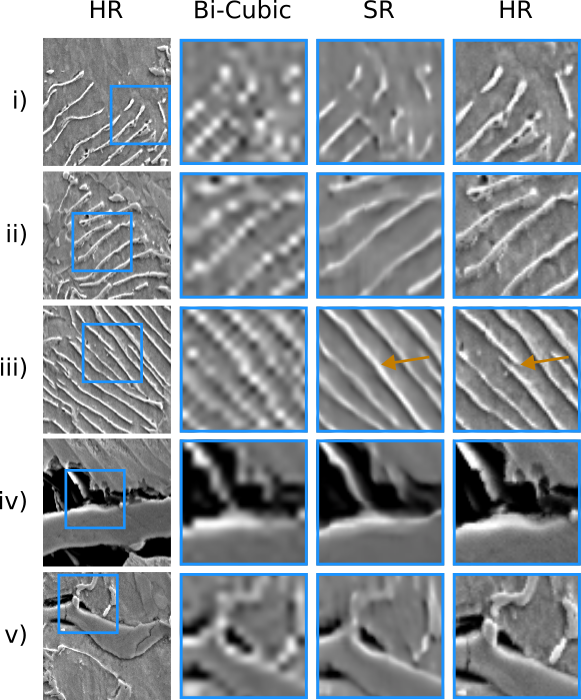}
    \caption{Different sections of the 16MnCrS5 microstructure shows in bi-cubic interpolation, super-resolution, and high-resolution ground truth. Each row depicts a larger area in high-resolution and one zoomed in region. i) Pearlite-Ferrite region, ii) Pearlite, iii) Pearlite, iv) MnS, v) MnS with surrounding ferrite.}
    \label{fig:mncrsphases}
\end{figure}

As direct application yielded overly bright images with artefacts, we investigate whether finetuning with a small dataset is sufficient to produce realistic images. For this we use a dataset consisting of $2,000$ low-resolution, high-resolution, reference images, which took about \SI{20}{\minute} to record. The resulting similarity metrics are displayed in \autoref{tab:16MnCrS5Metrics}. The trained network improves both in terms of PSNR and SSIM over Lanczos and bi-cubic interpolation. 
In \autoref{fig:mncrsphases} we present various microstructure examples. In the first three rows, different examples of pearlitic regions are shown. The network produces clearer cementite lamellae that closely resemble the high-resolution ground truth. In the third row an example of an erroneous network prediction is depicted. The network predicts that the centre lamella is one continuous lamella, while the ground truth image reveals, that at this spot the lamella is disjointed. In the last two rows two different MnS inclusions are shown.

\begin{table}[ht]
\centering
\caption{Similarity metrics for bi-cubic, Lanczos interpolation and super-resolution on the aligned 16MnCrS5 dataset.}
\label{tab:16MnCrS5Metrics}
\begin{tabular}{lccc}
\toprule
Metric &  Bi-Cubic & Lanczos & TTSR \\ \midrule
SSIM &  \SI{0.476}{} $\pm$ \SI{0.003}{}&\SI{0.467}{} $\pm$ \SI{0.003}{} & \SI{0.492}{} $\pm$ \SI{0.004}{} \\ 
PSNR &  \SI{18.4}{\decibel} $\pm$ \SI{0.1}{\decibel}&\SI{18.1}{\decibel} $\pm$ \SI{0.1}{\decibel} & \SI{19.1}{\decibel} $\pm$ \SI{0.1}{\decibel} \\ 
\bottomrule
\end{tabular}
\end{table}

\subsection{Time Efficiency in Direct Comparison}
\label{sec:time_efficiency}

Scanning an area with a quarter of the resolution, as performed in this work, reduces the required time by a factor of \SI{16}{}. Recording \SI{100}{\micro\meter} with a resolution of $4096\times 4096$ and a dwell time of \SI{32}{\micro\second} takes \SI{9}{\minute} while scanning the same area with a resolution of $1024 \times 1024$ takes about \SI{30}{\second}. Applying the TTSR algorithm for resolution enhancement takes an additional \SI{12}{\second} on a NVidia RTX3090 GPU. Since the imaging of larger areas in high-resolution is limited both by the field of view, as well as the working distance and lens configuration, large areas are usually imaged in patches. After the recording of each patch, the current patch can be enhanced by the network, while the microscope records the next patch, resulting, in no further overhead due to the application of the super-resolution algorithm.

\begin{figure}
    \centering
    \includegraphics[width=0.75\textwidth]{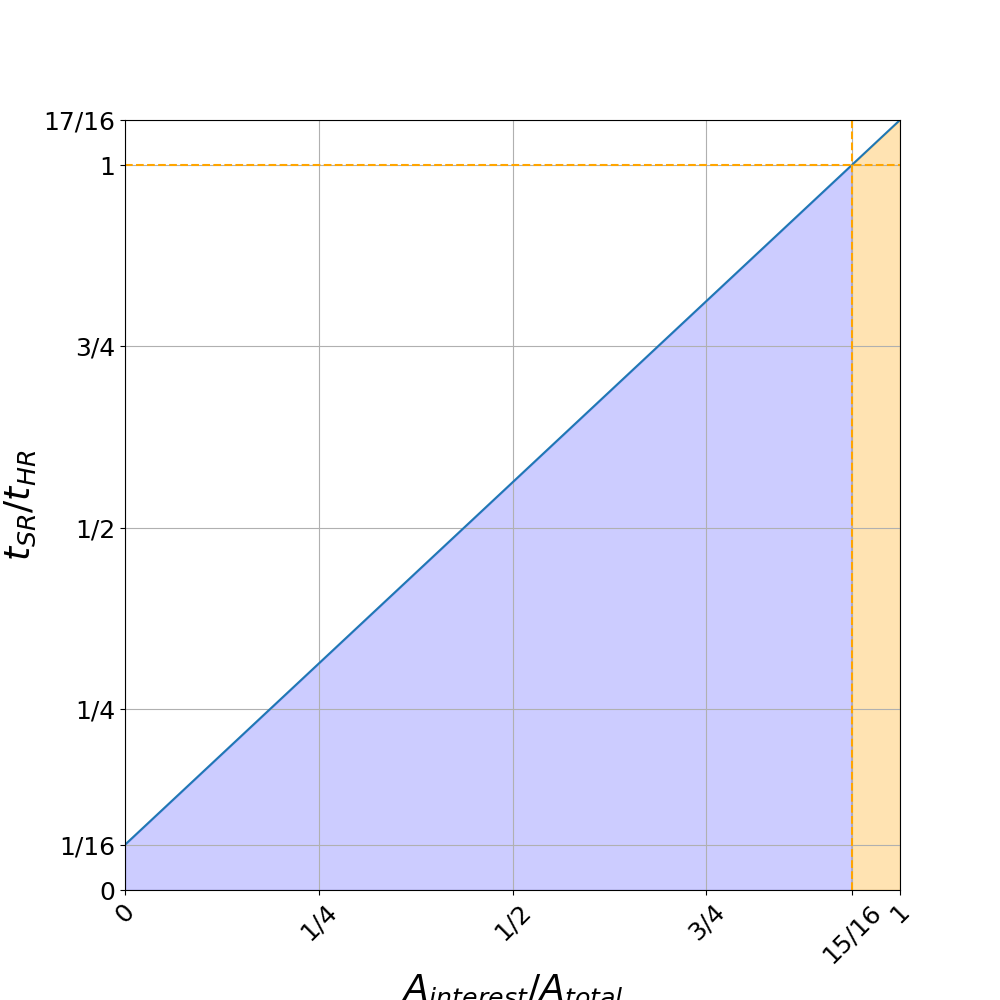}
    \caption{Ratio of time following a scan-enhance-rescan strategy and the time to perform a high-resolution scan plotted against the ratio of the area of interest and the total area. Marked in blue is the ratio interval, where our method is faster than a traditional large-area high-resolution scan, marked in orange, where performing a large-area high-resolution scan is faster than the presented method.}
    \label{fig:timeComparison}
\end{figure}

Assuming the application of this method is used in conjunction with re-scanning parts of the surface, as they have been identified as points-of-interest in the super-resolution image, we can estimate the time-saving per scan, given a trained super-resolution network as
\begin{equation}
    t_{SR}/t_{HR} = 1/16 + A_{interest}/A_{total}
\end{equation}
where $t_{SR}$ is the time required for a low-resolution scan and subsequent re-scanning of parts of the surface in a high-resolution, $t_{HR}$ is the time to scan the entire surface in a high-resolution, $A_{interest}$ is the area of the material that is of interest for later analysis, and $A_{total}$ is the total area to be imaged. Up to a ratio of $15/16$ the super-resolution based approach is faster than a large area high-resolution scan. This is depicted in \autoref{fig:timeComparison}.

\section{Discussion}

\subsection{Acceleration of Experiments}
\begin{figure}
    \centering
    \includegraphics[width=\textwidth]{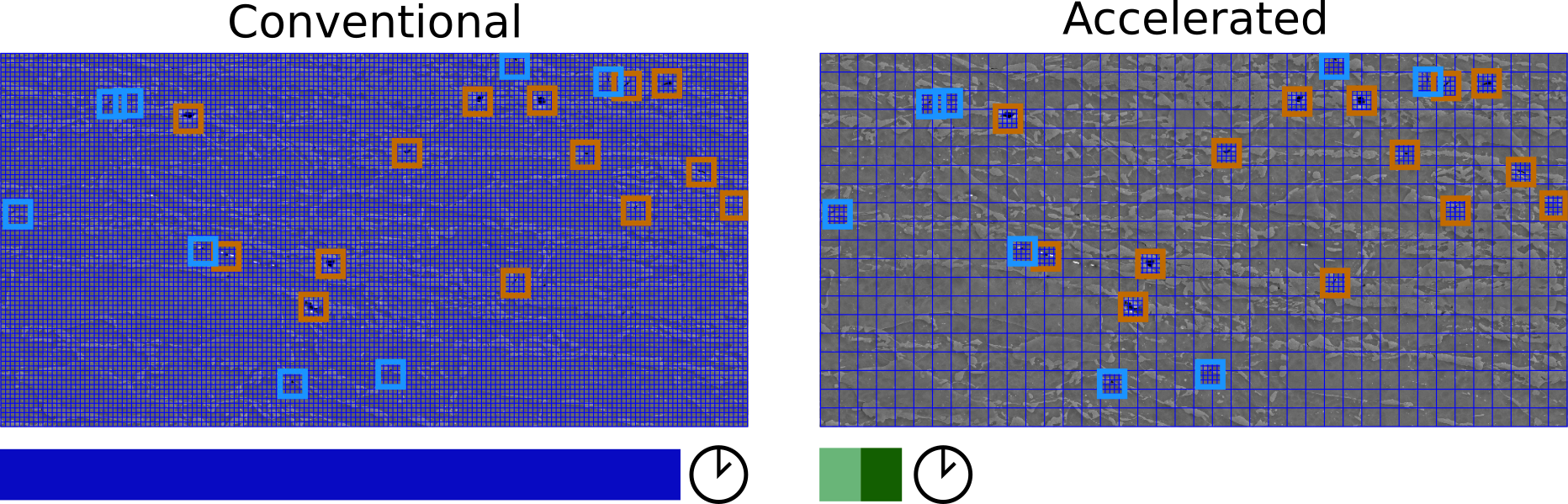}
    \caption{Illustration of time comparison between a conventional large-area high-resolution imaging and a super-resolution based accelerated imaging, employing a scan-enhance-rescan approach. Large-area high-resolution imaging images large parts of the surface not containing any relevant feature, while in the super-resolution based approach a first low-resolution scan of the surface identifies possible points-of-interest, which then can be re-scanned in a higher resolution. This results in the same information for later analysis. The bars under the images indicate the required time for the full process to finish. In the accelerated plot the time is divided over initial low-resolution scanning and additional high-resolution re-scanning of points of interest.}
    \label{fig:illustratedTimeComparison}
\end{figure}

Through the successful application of the method developed here, initial large area imaging can be significantly accelerated, here by a factor of $1/16$. Recording an area of \SI{1}{\square\milli\meter} at a resolution of \SI{32.5}{\nano\meter} per pixel with a dwell time of \SI{32}{\micro\second} would take about \SI{9}{\hour}, recording the same area with a quarter of the resolution in each dimension would only require about \SI{30}{\minute}. While the achieved similarity is high both visually as well as in terms of similarity metrics, differences still persist. Therefore, after the identification of points of interest, an additional re-scanning of these points in the required resolution for classification will be necessary. This additional step, will add on top of the low-resolution scanning, but as it is dependent on the ratio of the area of interest to the total area, the rarer studied events are, the higher the expected gain. We approximate that for $1000$ damage sites in a dual-phase steel panoramic image, we would require an additional time investment of \SI{30}{\minute}, resulting in a total time investment of one hour and a time-save of eight hours. This time comparison is visualised in \autoref{fig:illustratedTimeComparison}. In \cite{kusche_large-area_2019} the necessary time for the manual evaluation of $1000$ damage sites would require about \SI{5.5}{\hour}, while the methodology developed there, using neural networks for the evaluation would only require about one minute. The end-to-end evaluation using our approach combined with an automated evaluation of the recorded data would only require one hour contrasted to the traditional approach requiring \SI{14.5}{\hour}. Through this speed-up it becomes possible to analyse more material parameter configurations, even larger areas, more in-situ steps, more 3D slices in the same time.

\subsection{Application to Dual-Phase Steel Micrographs}
The network trained on the dual-phase steel dataset showed great visual similarity to the ground truth and a large improvement over bi-cubic interpolation. Generally, the sharp contrast of the different phases and voids is restored in the super-resolution images, while the interpolation images often produce blurry images. While these features are reconstructed with a high degree of visual similarity, predictions of fine grained sections can differ from the ground truth. On the other hand, features that are material specific and that can be reconstructed from low-resolution images, are reconstructed. For example a martensite crack, see \autoref{fig:largerMicrostructurePlot}, with a void pixel distribution as present in the low-resolution image, probably depicts a complete martensite crack, is reconstructed accordingly. The crack in the interpolated image appears to have small bridges still connecting the two martensite islands. In this case the network has probably learned, that such a configuration does not occur in the dataset and reconstructed the crack correctly. There are other features in the dataset where such a unique mapping between low- and high-resolution images is not possible, as for small martensite grains. In these cases the network has difficulty in reconstructing the exact shape of the high-resolution ground truth, and will make a conservative prediction. 

The visual improvement over interpolation is partly reflected in terms of similarity metrics, but the magnitude of the improvement seems to not be fully represented. We attribute this to the different philosophies followed by interpolation and super-resolution. In terms of metrics, interpolation is a good conservative approximation of the high-resolution images, where coinciding pixel, disregarding noise, have the same value. In between the recorded low-resolution pixel now a smooth function is approximated, that on average will lie close to the high-resolution ground truth values. Super-resolution on the other hand optimises additional metrics, next to the reconstruction loss, which is closely related to PSNR. Its predictions therefore follow a less conservative approach which produces sharper features in contrast to the smooth interpolation features. As these sharp features correspond to fast changes in the pixel values, small deviations in position can strongly negatively impact the similarity metrics.

Furthermore, we applied the trained network to a larger micrograph in order to test, whether the enhanced smaller patches can be assembled to give researchers more contextual information and test how the network performs on data on the same material in slightly different conditions. The assembled larger micrograph does not show any stitching artefacts at the edges of the predictions. The predictions of the network are again close to the high-resolution ground truth and appear to be more smoothed out compared to the predictions on the patches directly recorded in the networks input size. This can also be seen in terms of similarity metrics, while our network achieves a PSNR score of \SI{25.96}{\decibel} $\pm$ \SI{0.04}{} on the first dataset, it only achieves a PSNR score of \SI{22.2}{\decibel} $\pm$ \SI{0.1}{\decibel}. We attribute this to a difference in data fidelity. While the first dataset was recorded specifically for the network to be trained on, using a smaller window size, thereby minimising effects such as drift build up, the second dataset was recorded in the largest available window size present at the electron microscope. As features do not align between the low-resolution and high-resolution counter pairs in the second dataset, a re-alignment pre-processing step is necessary. While this step re-aligns features, noise can lead to small mis-alignments, thereby leading to a smaller similarity measure. This highlights the necessity of a well curated dataset especially during training. 

\subsection{Transfer to 16MnCrS5 Steel Micrographs}
The direct application of the network trained on the dual-phase steel dataset resulted in overly bright images, even after re-scaling of brightness and contrast. Furthermore, stitching of the predicted patches to a larger image resulted in stitching artefacts present at the borders of the prediction. We attribute this to the difference in brightness and contrast during recording as well as the significantly different microstructure of the 16MnCrS5 steel. Unifying the imaging conditions between different materials could mitigate this issue to some extent in future works. 

Training with a small dataset consisting of $2,000$ data points, improved the performance of the network, both in terms of similarity metrics as well as visual similarity. Here lies the highest possible achievable gain for a performance improvement. A network trained on a wider range of materials and imaging conditions, might serve as a stronger foundation model, that requires only a small dataset to fine tune on the material at hand.

\section{Conclusion}

In this work we successfully applied a deep learning reference-based super-resolution approach to scanning electron micrographs of both a dual-phase steel, as well as 16MnCrS5 steel surfaces. We were able to increase the resolution by a factor of four while performing better than classical interpolation approaches, both in terms of PSNR and SSIM metrics, as well as in direct visual comparison. Furthermore the following conclusions can be drawn:

\begin{itemize}
    \item Using the network trained here for resolution enhancement and the presented scan-enhance-rescan strategy imaging in the scanning electron microscopy can be accelerated up to a factor of $16$, where the specific acceleration depends on the ratio of area of interest to the total area. The smaller this ratio, the larger the expected acceleration, making this method particularly useful for the study of rare events, such as the study of damage mechanisms.
    \item While similarity metrics can quantify the similarity between predictions and ground truth images, they can miss important image characteristics, like sharp phase contrasts. Predicting these features can negatively affect similarity metrics, when these features are off by a small amount, as they correspond to sharp changes in pixel values.
\end{itemize}


\newpage

\section*{Acknowledgements}
The authors express their gratitude to the Deutsche Forschungsgemeinschaft (DFG) for financial support in context of the Collaborative Research Centre CRC/Transregio 188/2 ‘‘Damage Controlled Forming Processes”, project T02 and B02, project no. 278868966.
Calculations were performed with computing resources granted by RWTH Aachen University.

\section*{Data and Code Availability}
Will be made available upon publication and upon request during peer-review.

\bibliographystyle{ieeetr}
\bibliography{
    bibliography/MyLibrary.bib
}

\begin{appendices}
\renewcommand\thefigure{\thesection.\arabic{figure}} 
\setcounter{figure}{0}

\section{Appendix}
\subsection{More Dual-Phase Steel Microstructure Examples}
\label{sec:moreExamples}
\begin{figure}
    \centering
    \includegraphics[width=0.8\textwidth]{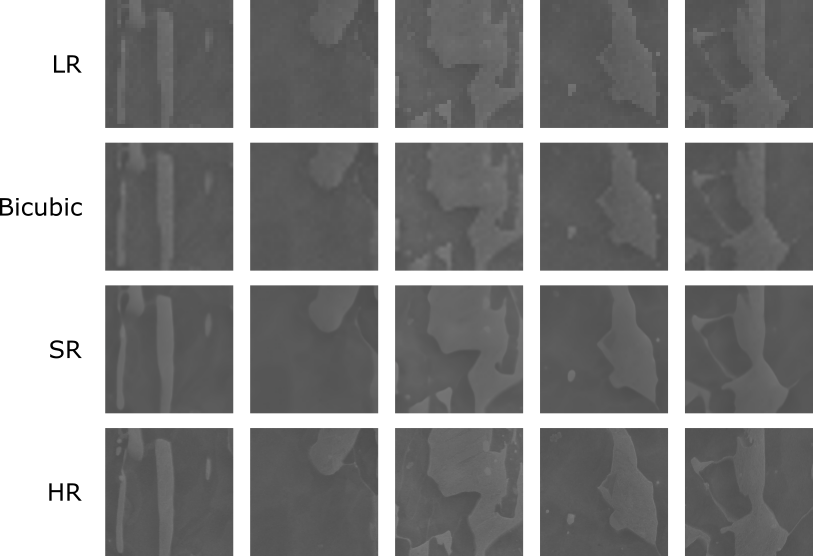}
    \caption{Parts of the microstructure in low-resolution, bicubic interpolation, super resolution, and high-resolution.}
    \label{fig:dpMicrostructure}
\end{figure}
\begin{figure}
    \centering
    \includegraphics[width=0.8\textwidth]{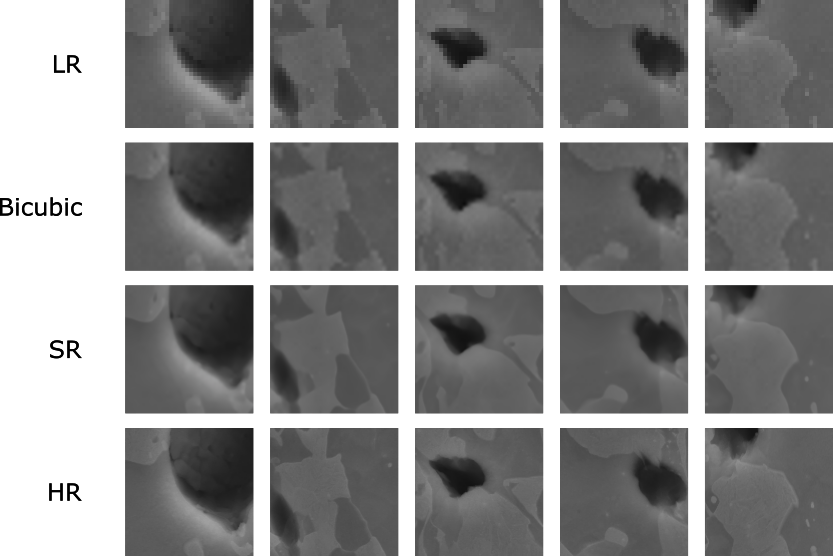}
    \caption{Inclusions from the test dataset in low-resolution, bicubic interpolation, super resolution, and high-resolution.}
    \label{fig:dpmInclusions}
\end{figure}
\begin{figure}
    \centering
    \includegraphics[width=0.8\textwidth]{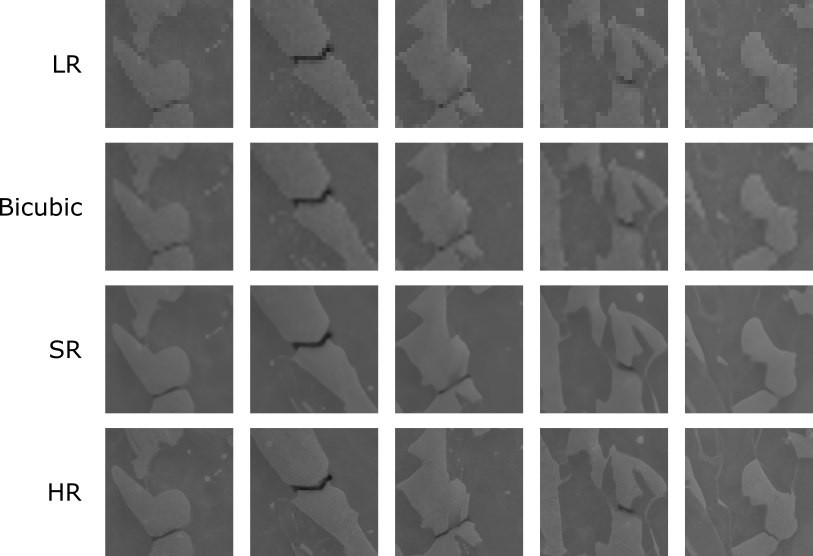}
    \caption{Martensite cracks from the test dataset in low-resolution, bicubic interpolation, super resolution, and high-resolution.}
    \label{fig:dpMartensiteCracks}
\end{figure}
\begin{figure}
    \centering
    \includegraphics[width=0.8\textwidth]{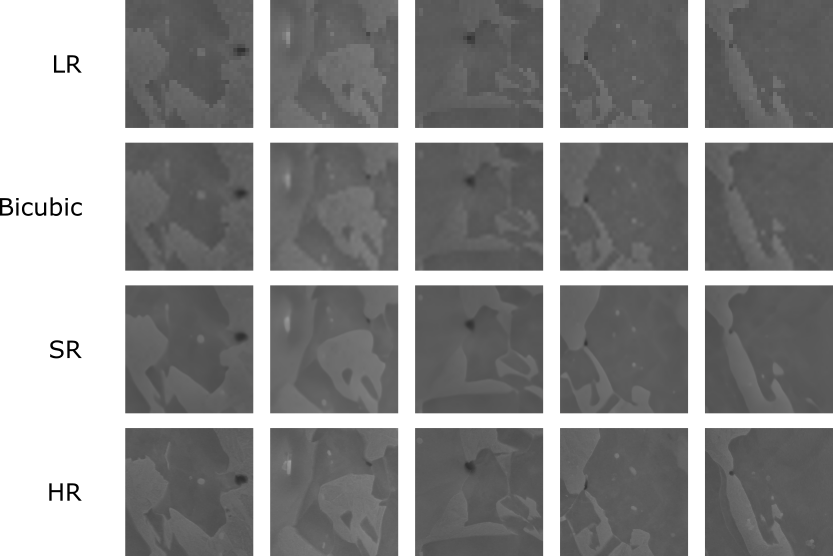}
    \caption{Interface decohesions from the test dataset in low-resolution, bicubic interpolation, super resolution, and high-resolution.}
    \label{fig:dpInterfaceDecohesions}
\end{figure}
\begin{figure}
    \centering
    \includegraphics[width=0.8\textwidth]{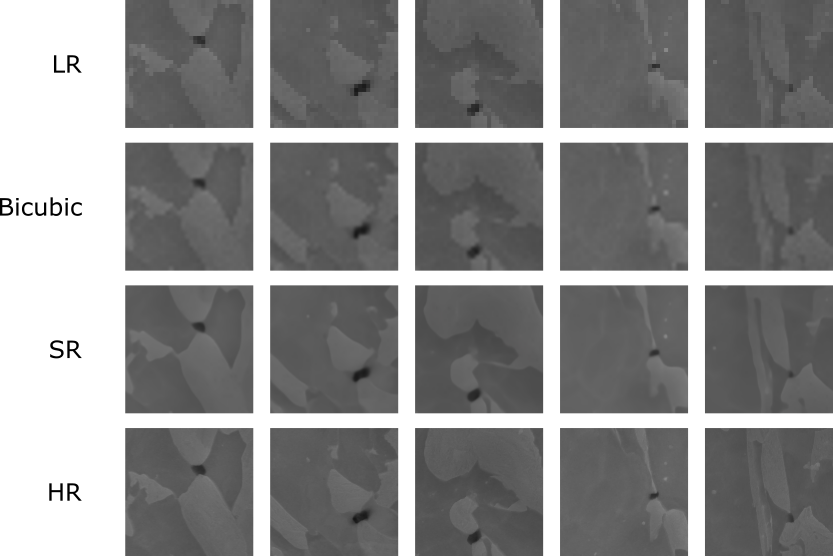}
    \caption{Martensite cracks from the test dataset in low-resolution, bicubic interpolation, super resolution, and high-resolution.}
    \label{fig:dpNotchEffect}
\end{figure}
\begin{figure}
    \centering
    \includegraphics[width=0.8\textwidth]{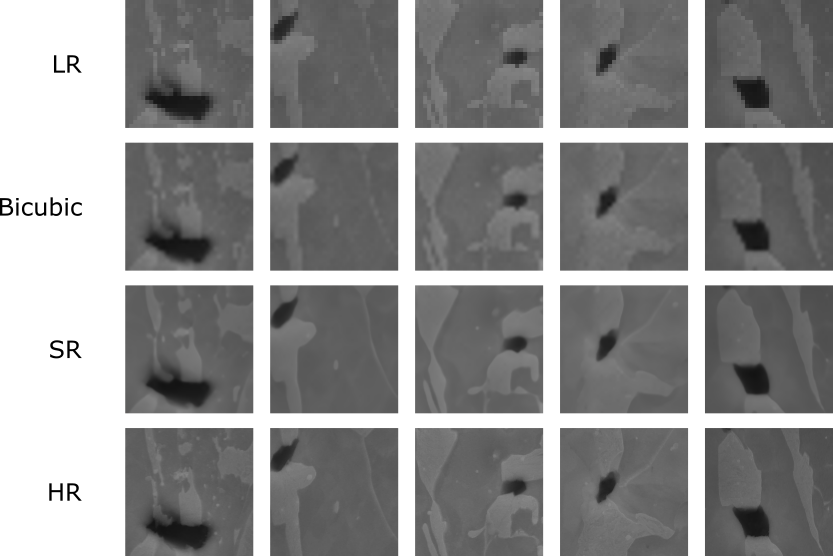}
    \caption{Martensite cracks from the test dataset in low-resolution, bicubic interpolation, super resolution, and high-resolution.}
    \label{fig:dpEvolved}
\end{figure}
\begin{figure}
    \centering
    \includegraphics[width=0.5\textwidth]{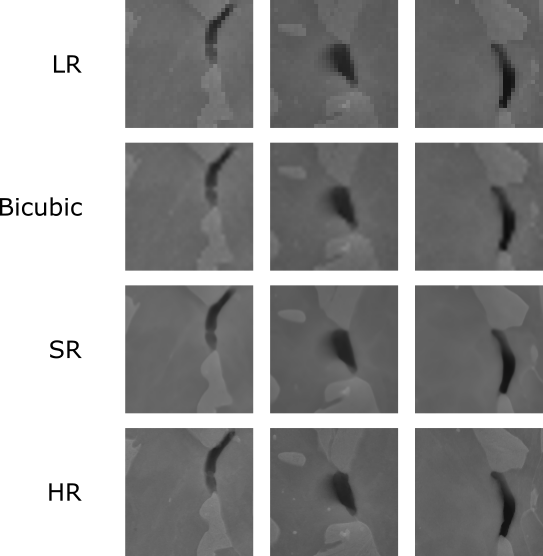}
    \caption{Martensite cracks from the test dataset in low-resolution, bicubic interpolation, super resolution, and high-resolution.}
    \label{fig:dpBoundaryDecohesion}
\end{figure}

\end{appendices}
\end{document}